# A Novel Interconnect Camouflaging Technique using Transistor Threshold Voltage

Jae-Won Jang[1] and Swaroop Ghosh[2]
School of Electrical Engineering and Computer Science
Pennsylvania State University, University Park, PA-16801, USA
{jxj328[1] and szg212[2]}@psu.edu

*Abstract*—Semiconductor supply chain is increasingly getting exposed to variety of security attacks such as Trojan insertion, cloning, counterfeiting, reverse engineering (RE) and piracy of Intellectual Property (IP) due to involvement of untrusted parties. Camouflaging of gates has been proposed to hide the functionality of gates. However, gate camouflaging is associated with significant area, power and delay overhead. In this paper, we propose camouflaging of interconnects using multiplexers (muxes) to protect the IP. A transistor threshold voltage-defined pass transistor mux is proposed to prevent its reverse engineering since transistor threshold voltage is opaque to the adversary. The proposed mux with more than one input, hides the original connectivity of the net. The camouflaged design operates at nominal voltage and obeys conventional reliability limits. A small fraction of nets can be camouflaged to increase the RE effort extremely high while keeping the overhead low. We propose controllability, observability and random net selection strategy for camouflaging. Simulation results indicate 15-33% area, 25-44% delay and 14-29% power overhead when 5-15% nets are camouflaged using the proposed 2:1 mux. By increasing the mux size to 4:1, 8:1, and 16:1, the RE effort can be further improved with small area, delay, and power penalty.

## I. INTRODUCTION

Reverse Engineering (RE) of an Intellectual Property (IP) [1-2] is a process of identifying its design, functionality, and structure. In the RE method, the adversary de-layers the IC, determines the gate functionalities and their connectivity information, and, reconstructs the netlist (Fig. 1). This technique has been originally used by industries with the mindset of gathering information on its competitors, to confirm the functionality of their own design, and to ensure the legitimacy of circuits from the piracy. However, the advanced adversaries can exploit this technique with an ill intention to steal and pirate a design to illegally sell in the black market. Split manufacturing [1] technique has been proposed to make the IC fabrication more secure and robust against RE while simultaneously alleviating the cost of owning a trusted foundry. Split manufacturing technique separately manufactures the front-end (transistors) in an untrusted foundry whereas the back-end (interconnect) is manufactured in trusted facility. This makes the RE and Trojan insertion more challenging for the adversary since the connectivity information is hidden in the untrusted foundry. Furthermore, since the front-end fabrication cost is higher than the back-end, the cost benefit of outsourcing the fabrication is still preserved without increasing the security risks. Although this technique is effective in preventing RE, it can be susceptible to yield loss during stacking due to via misalignment. Furthermore, it still requires trusted foundry and costly assembly process.

Camouflaging of gates have been proposed [10-11] to affordably hide the logic functionality and make the RE economically non-profitable or extremely difficult. The primary objective of gate camouflaging is to hide the functionality of *few chosen gates* (since camouflaged gates are typically area, delay and power intensive) to increase RE effort of adversary while keeping power, performance and area overhead minimal. The camouflaged gates can assume functionalities such as AND, OR, XOR, etc. Although the exact gate functionality is hidden, the adversary can still create a partial netlist with other known gates and go through guess-and-validate process to RE the missing gate functionality. This is achieved by making a guess about the gate function, finding test patterns to confirm the guess, and then applying these patterns to both a partial netlist and a golden chip. If the outputs match, then the guess is correct; else the adversary guesses a new gate functionality and repeats the procedures. This process is shown in Fig. 2(a). The RE effort is also shown which involves the time needed to identify all camouflaged gate functionalities.

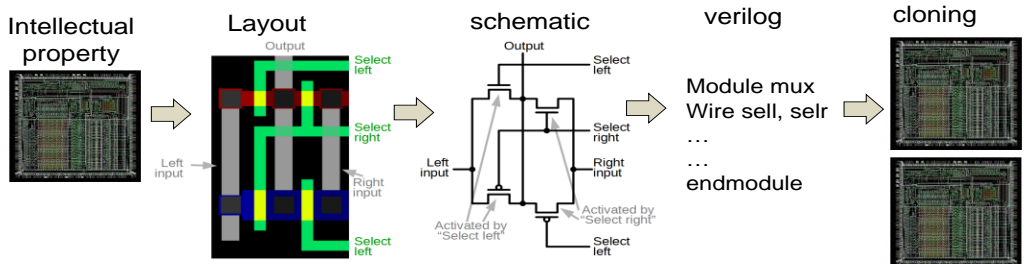

Fig. 1. Reverse engineering of IP: the chip is de-layered to identify the gate functionality and their connectivity which is used to reconstruct the schematic and netlist. The objective is to clone the design.

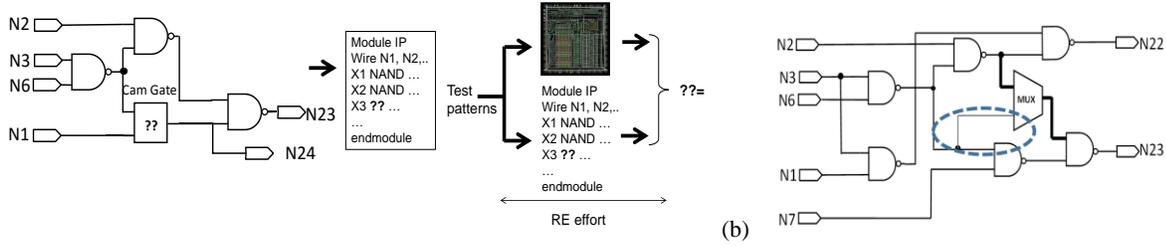

(a)                                                                                    (b)

Fig. 2. (a) When a gate is camouflaged, the adversary extracts the partial netlist, guesses the missing gate functionality ("??") and applies specific test pattern to match the output against actual chip to confirm the guess. The RE effort is the time invested by adversary to find appropriate test pattern and identify the camouflaged gate functionality; and, (b) the concept of interconnect camouflaging using mux. The real connection is shown using thick lines. The RE effort will involve guessing a connection and validating it by running test patterns.

It has previously been shown that careful camouflaging of ~10-40% gates can increase the RE effort significantly [8]. Dummy contact [1-2], programmable standard cell [3], and filler cell [4] based camouflaging have been proposed in past for Integrated Circuit (IC). For sequential circuits, additional logic (black) states are introduced in the finite state machine [9], which allow the design to reach a valid state only using the correct key. In combinational logic, XOR / XNOR gates are introduced to conceal the functionality [13-14]. A new gate camouflaging technique based on the transistor threshold voltage ($V_T$) programmable switch that turns ON/OFF based on $V_T$ assertion is also proposed for camouflaging [12][18].

In this paper, we introduce a novel interconnect camouflaging technique to hide the connectivity information between gates in contrary to the widespread gate camouflaging methodologies. This technique is conceptually similar to split manufacturing with two major distinctions: (i) the proposed interconnect camouflaging technique does not require the splitting of layers between trusted and untrusted foundry; and (ii) only few selected nets are camouflaged. By carefully selecting the nets based on the prioritized metrics (which will be later discussed), we can maximize the RE effort. It has been shown that threshold voltage could be reverse engineered by inspecting the transistor doping using Focused Ion Beam (FIB) and Scanning Electron Microscopy (SEM) [24], this process is expected to be very expensive. Furthermore, since ICs contain billions of transistors, identifying dopant levels of a fraction of transistors could be tedious. Therefore, the IP could still be protected from low-cost optical RE.

The proposed camouflaging technique is achieved by inserting muxes in the design as illustrated in Fig. 2(b) using a C17 circuit as an example (from ISCAS85 benchmark [21]). The original connection is shown in thick black lines whereas the dummy connection is shown in light grey (highlighted in dashed-circle). If the adversary does not know the multiplexer select signal, he will resort to brute-force method using trial-and-error. This, in turn, will increase the RE effort. The RE effort could be further increased by adding more fake signals and using a N:1 mux. Since mux design is low-overhead in comparison to the gate camouflaging, the proposed

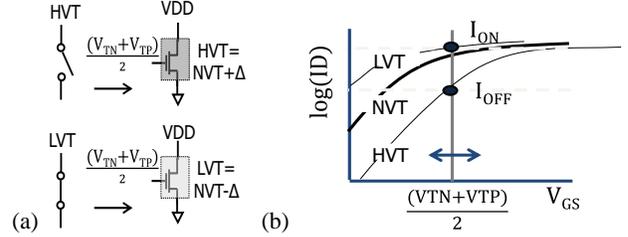

Fig. 3 (a) $V_T$ programmable switch. HVT: OFF, LVT: ON. PMOS switch works similarly; and (b) cartoon of I-V curves of NVT, HVT and LV transistors. The $I_{ON}$ and $I_{OFF}$ depends on the LVT and HVT values as well as on gate voltage biasing.

interconnect camouflaging technique is light-weight while being effective.

The disadvantage of the above method is the requirement of a select signal for muxes. To resolve this issue, we propose a novel multiplexer design based on transistor $V_T$–defined switches proposed in [18]. This eliminates the need of select signal and leaves no layout trace for the adversary to RE. In the proposed mux, the selected (real) path will connect to the output through low $V_T$ transistors whereas the fake paths will be disconnected from the output using high $V_T$ transistors. As the transistors $V_T$ are implemented by changing channel doping concentration during the manufacturing process, this information is opaque to the adversary forcing him to resort to RE-intensive trial-and-error approach.

Furthermore, **$V_T$ modulation** (mixing normal $V_T$ (NVT), high $V_T$ (HVT) and low $V_T$ (LVT) transistors in a circuit) is a well-known technique that is extensively utilized [7-8] in semiconductor industries for the trade-off between power, performance and robustness. Therefore, the proposed methodology does not require any additional process steps and can be extended to N:1 mux to increase RE effort of the adversary.

It is important to note that the proposed camouflaging is robust against de-camouflaging solution proposed in [25] since it does not involve camouflaging of gates. Instead we camouflage the net connections which requires a new de-camouflaging methodology. Similar concept of $V_T$-based camouflaging has been proposed in [12]. However, it proposes dynamic gates which differs from the proposed interconnect camouflaging technique. We integrate the $V_T$-defined muxes in different ISCAS benchmarks to evaluate the effectiveness of interconnect camouflaging. *To the best of our knowledge, this is the first effort towards camouflaging the interconnects using $V_T$-defined multiplexers*. In particular, we make following contributions in this paper:

- We propose an interconnect camouflaging technique for protection of IP. The proposed interconnect camouflaging is low-overhead compared to camouflaged gates and can potentially eliminate the need of split manufacturing.
- We propose a $V_T$ defined 2:1 and N:1 (where N=4, 8 and 16) mux to obfuscate the nets. The muxes are optimized for optimum area, delay and power overhead.
- We propose interconnect selection procedure and analysis framework to quantify area, power and delay overheads, and estimate RE effort.
- We conduct thorough analysis of the proposed gate at system level using interconnect selection algorithms and RE effort metrics.

The rest of the paper is organized as follows. In Section II, we describe the design and analysis of the proposed $V_T$-defined mux. Proposed interconnect obfuscation approach is presented in Section III. The metrics for camouflaging and simulation results are described in Section IV. Conclusions are drawn in Section V.

## II. THRESHOLD VOLTAGE DEFINED MULTIPLEXER

In this section, we present the concept of $V_T$-defined switch. This is followed by the proposed $V_T$-defined mux and key design requirements. Next, we present the mux design space exploration to achieve desired robustness and performance.

### A. Threshold Voltage defined Switch

We use the programmable switch proposed in [18] that turns ON/OFF based on $V_T$ asserted on it. In this work, the switch is optimized to suit the mux application. The switch is realized by using conventional NMOS and PMOS transistors with the gate biased at the mid-point between nominal NMOS and PMOS threshold voltages i.e., $V_{SN}= 0.5(V_{TN}+V_{TP})$. Therefore, the switch conducts when low $V_T$ (LVT) is assigned during manufacturing. This is because $V_{GS}= 0.5(V_{TN}+V_{TP}) >$ LVT. The switch stops conducting when high $V_T$ (HVT) is assigned since $V_{GS}=0.5(V_{TN}+V_{TP}) <$ HVT. This is depicted in Fig. 3(a) for NMOS transistor. The cartoon of transistor I-V curves for NVT, LVT and HVT transistor is shown in Fig. 3(b). The $I_{ON}$ and $I_{OFF}$ that can be obtained by assigning LVT and HVT is also shown. A good $V_T$ defined switch should offer high ON current and low OFF current. The gate voltage, HVT, LVT values and transistor sizes are tuned to maximize the $I_{ON}/I_{OFF}$ ratio. For NMOS-switch, higher HVT values and lower gate voltage is good for $I_{OFF}$ (leakage) whereas lower LVT and higher gate voltage is good for $I_{ON}$ (performance). Vice-versa is true for PMOS-switch. The switch optimization in presence of these conflicting requirements is described in Section IIC.

### B. Multiplexer Design and Challenges

The $V_T$-defined switch [18] is optimized to suit the mux application. In the proposed mux, the real path contains LVT pass transistor and the fake paths contain HVT pass transistor (Fig. 4(a)). This eliminates the need of a mux select input as $V_T$ value inherently determines the input selection. Since an NMOS transistor cannot pass a strong input '1', we incorporate a level restoring weak HVT PMOS transistor (highlighted with dashed-circle in Fig. 4(a)) to pull the NMOS pass transistor output to full-rail. The level restoring transistor helps full voltage swing of the degraded input and improves the low-to-high transition. Furthermore, it eliminates the static current from the output inverter. The sizing of this level restoring PMOS transistor is done carefully so that it does not contend with the mux inputs. The alternative design technique to avoid level restoring transistor is to use full transmission gates (with NMOS and PMOS in parallel as shown in Fig. 4(b)). This method will allow both strong input '0' and '1' to be passed through the muxes, but incurs significant power,

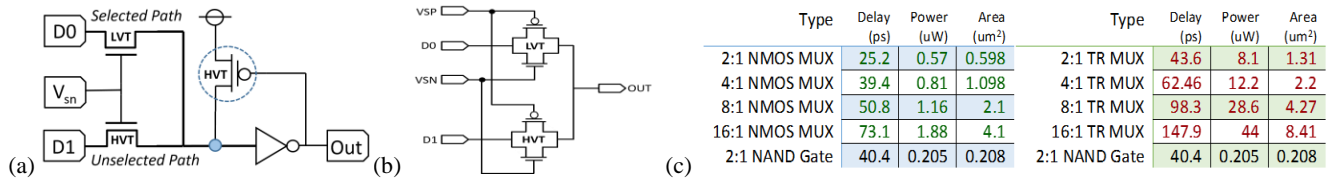

| Type | Delay (ps) | Power (uW) | Area (um²) | Type | Delay (ps) | Power (uW) | Area (um²) |
|---|---|---|---|---|---|---|---|
| 2:1 NMOS MUX | 25.2 | 0.57 | 0.598 | 2:1 TR MUX | 43.6 | 8.1 | 1.31 |
| 4:1 NMOS MUX | 39.4 | 0.81 | 1.098 | 4:1 TR MUX | 62.46 | 12.2 | 2.2 |
| 8:1 NMOS MUX | 50.8 | 1.16 | 2.1 | 8:1 TR MUX | 98.3 | 28.6 | 4.27 |
| 16:1 NMOS MUX | 73.1 | 1.88 | 4.1 | 16:1 TR MUX | 147.9 | 44 | 8.41 |
| 2:1 NAND Gate | 40.4 | 0.205 | 0.208 | 2:1 NAND Gate | 40.4 | 0.205 | 0.208 |

Fig. 4 (a) The proposed pass transistor NMOS-only based 2:1 mux; (b) the Transmission (TR) gate based 2:1 mux; and (c) attributes of the proposed N:1 mux for NMOS-only mux and TR-gate mux delay, and size overhead due to requirement of large PMOS transistors. As Fig. 4(c) shows the comparison result, transmission gate design increases area and power overhead especially for wide input mux designs. The pass transistor NMOS-only mux logic allows the proposed design to be compact without incurring significant overhead. The 2:1 NMOS mux is comparable to 2-input minimum sized NAND gate in terms of delay.

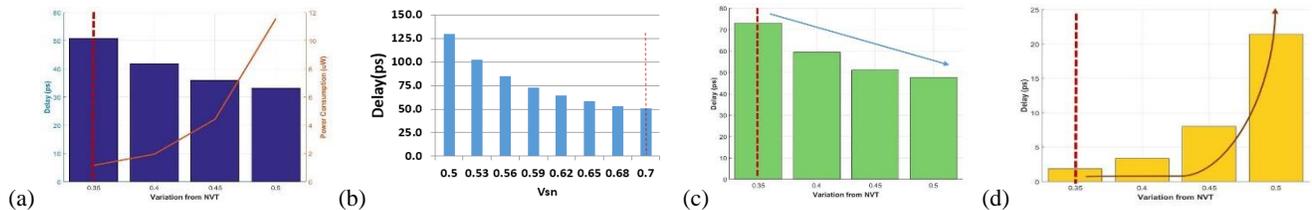

Fig. 5. Selection of $V_{SN}$ and offset from NVT: (a) 8:1 mux delay and leakage vs offset; (b) 8:1 mux delay vs $V_{SN}$; (c) 16:1 mux delay vs offset; and, (d) 16:1 mux leakage vs offset. The optimal choice of offset is also shown by dashed lines.

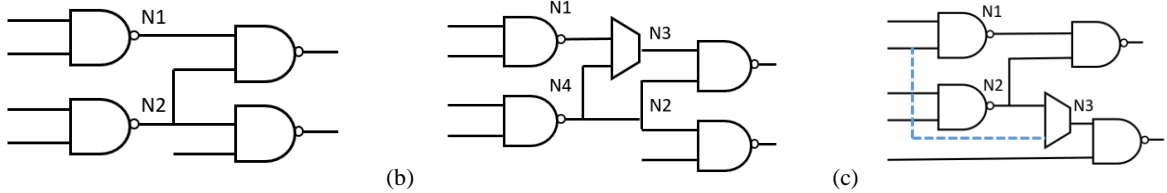

Fig. 6. Example of qualified nets selection: (a) original circuit; (b) single-fan-out net which is selected for mux insertion. Since N1 cannot float, the adversary can easily guess that N1 connects to N3. We disqualify such nets to prevent mux insertion; and (c) multi-fan-out net N2 is selected for mux insertion. The adversary cannot figure the connection between N2 and N3. Such nets are qualified to enable selection.

### C. Design Space Exploration

For the design space exploration, we have used Nangate 45nm technology [19]. The LVT and HVT values are chosen by determining their offset ($\Delta$ in Fig. 3(a)) from NVT value. For example, if the NVT of NMOS transistor is 0.62V, an offset of 0.1V (i.e., $\Delta = 0.1V$) means that the LVT is 0.52V and HVT is 0.72V. For simulations, we have used 8:1 mux and 16:1 mux circuit. We sweep both offset and gate voltages ($V_{SN}$) and calculate the delay and leakage power. The offset voltage ($\Delta$) is swept from 0.30V to 0.45V in steps of 0.05V for NMOS as well as PMOS. The switch gate voltage $V_{SN}$ is swept from 0.1V to 0.5V in steps of 0.05V. Fig. 5(a) shows the delay and leakage power values with offset and Fig. 5(b) shows the delay when $V_{SN}$ are varied. From these two plots, we choose the optimum values of $\Delta$ (= 0.35V), and $V_{SN}$ (=0.7V) which are used for simulating 2:1, 4:1, 8:1, and 16:1 muxes. A similar trade-off study is conducted between the delay and leakage power for 16:1 mux (Fig. 5(c)-(d)).

It is noteworthy that LVT and HVT in a process technology is optimized based on factors such as, leakage and performance of combinational logic. Therefore, it is likely that the proposed camouflaging will end up using the pre-defined HVT and LVT values. However, if the optimization option is made available to the camouflaging designer, then security could be factored in along with leakage and performance to decide the optimal values of HVT and LVT as described in this Section. Note that $V_{SP}$ is not shown in the exploration since we are not using $V_T$-defined PMOS transistor in the proposed mux.

## III. THRESHOLD-DEFINED CAMOUFLAGING METHODOLOGY

In this section, we present methodology to identify interconnects that can be obfuscated using $V_T$-defined multiplexers.

### A. Selection of Qualified Nets

```
Interconnect Camouflaging Algorithm
Input:  - Synthesized Netlist (ISCAS85)
        - Input value of nets replacement percentage
        - Sort priority selection input
Output: - Newly generated camouflaged netlist
Start
if input = true then
  1 > Create graph (G)
  2 > Traverse G; compute and record controllability
      (CC) and observability (Obs) of nets
  3 > Upon passing through the output net, assign
      metric M = CC0+CC1+Obs along with fan-out
      (FO) of the net
  4 > Identify / Assign qualified nets based on step 3
  5 > Sort in descending order based on the metric M
  6 > Select Dummy / Fake nets based on the priority
      of CC0, CC1, Obs, and FO parameters
  7 > Muxes are inserted in top x% (x is the
      percentage of nets to be camouflaged)
  8 > Generate a new camouflaged netlist
else
  1 > End simulation
End
```

Fig. 7. Pseudocode of the net selection methodology

To maximize the RE effort, it is critical to camouflage the nets that cannot be reverse engineered through simple intuition. For example, if net N1 (which is a single fan-out net) in Fig. 6(a) is camouflaged using a mux as shown in Fig. 6(b), then reverse engineering becomes straightforward. This is true since N1 cannot float in a valid design. This leaves the adversary to conclude that N1 and N2 are connected without running any simulation. We discard such single fan-out nets from the selection algorithm. In contrary, if a multi-fan-outs net such as N2 is selected for mux insertion, then the adversary cannot figure out the connection between N2 and N3 (Fig. 6(c)). Such nets are considered qualified nets in the proposed camouflaging procedure.

### B. Net Selection Methodology

The design objective is to identify interconnects based on quantifiable values to maximize RE effort. We first compute the controllability (CC) and observability (Obs) values for every net and its number of fan-outs in a circuit. The '0' and '1' controllability (CC0 and CC1) and observability values provides a relative difficulty of controlling and observing a logic signal of a particular net. By selecting the net with low

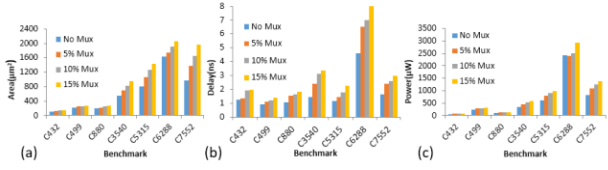

Fig. 8. Area, delay, power values of different benchmarks based on the percentage replacements (5 to 15%) of nets by 2-1 mux; (a) area; (b) delay; and (c) power of net selection methodology.

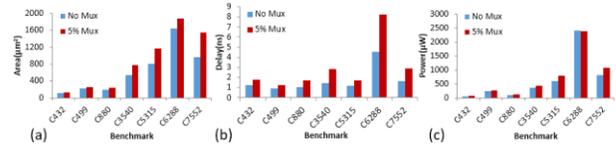

Fig. 9. Area, delay, power values of different benchmarks based on the percentage replacements (5%) of nets by 4-1 mux; (a) area; (b) delay; and (c) power of net selection methodology.

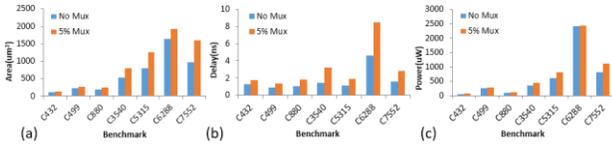

Fig. 10. Area, delay, power values of different benchmarks based on the percentage replacements (5%) of nets by 8-1 mux; (a) area; (b) delay; and (c) power of net selection methodology.

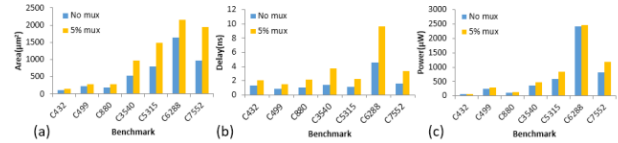

Fig. 11. Area, delay, power values of different benchmarks based on the percentage replacements (5%) of nets by 16-1 mux; (a) area; (b) delay; and (c) power of net selection methodology.

CC0, CC1 and Obs values, it is possible to increase the RE effort of adversaries. Note that the controllability and observability of the net is assigned the same value as the controllability and observability of the gate that is driving the net. For the nets with fan-outs, the controllability and observability is propagated to all fan-out nets.

Fig. 7 displays the proposed gate selection heuristic which is implemented in C++ and tested using HSPICE simulation. The heuristic imports Verilog benchmarks and finds controllability / observability values of the gates (step 2) and then assigns these values to the output nets and their fan-outs (step 3). The number of fan-outs of the nets is also calculated and assigned to the corresponding parent nets (step 4). Upon obtaining these parameters, we sort the output nets in descending order based on CC0+CC1+Obs value (step 5). We prioritize the number of fan-outs over the CC / Obs values since the higher number of fan-outs increases a circuit's RE effort. When nets are sorted, we select dummy / fake nets based on the priority of CC0, CC1, Obs, and fan-outs parameters (step 6). By selecting the fake nets that are difficult to control and observe, we can further improve RE effort. Afterwards, muxes are inserted in top x% (where x is target percentage of nets that needs to be camouflaged) of the nets (step 7). Upon finishing the mux insertion, new camouflaged netlist is generated (step 8). This netlist is used as an input for the Synopsys Design Compiler to perform synthesis and to evaluate the overall design in terms of area overhead, propagation delay, and power consumption compared with the original ISCAS85 benchmarks.

In addition to the above technique, we also select nets randomly for camouflaging. The fake nets are also randomly selected and the RE effort and overheads are compared with respect to the controllability/observability based selection methodology.

## IV. SIMULATION RESULTS

In this section, we evaluate the design overhead using Synopsis Design Compiler for ISCAS85 benchmarks [21]. Since $V_T$-defined muxes are not included in standard cell library, we have created a liberty file of the proposed muxes with values characterized using HSPICE simulation.

Fig. 8(a-c) shows area, delay, and power of benchmarks replaced with 2:1 muxes for 5%, 10% and 15% camouflaging using the controllability/observability based net selection methodology. Compared to the original ("No Mux") design, the average overhead is found to be 15% (area), 25% (delay) and 14% (power) for 5% camouflaging. The values for 10%

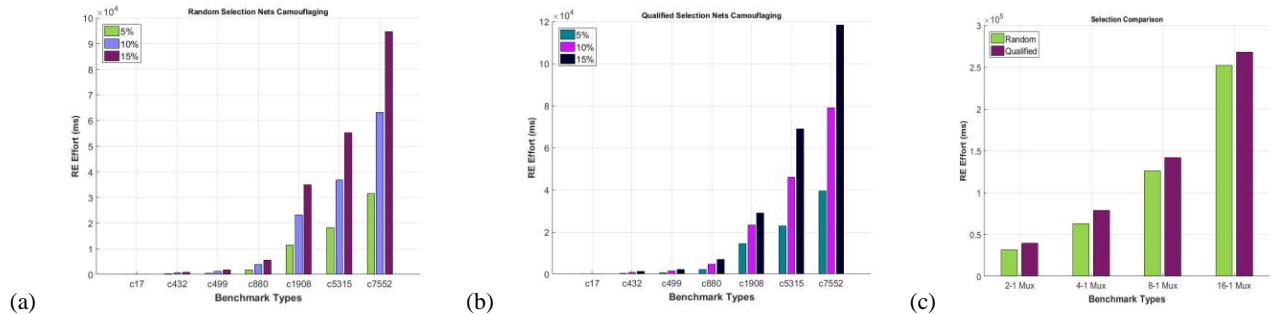

Fig. 12. 2:1 Mux RE effort based on the % camouflaging: (a) based on random selection of nets; (b) based on controllability and observability of nets; and (c) comparison for different types of muxes for 5/10/15% camouflaging on c7552 benchmark.

Table 1. Average overhead percentage of N:1 mux for 5% camouflaging of nets

| Gate | Area | Delay | Power |
|---|---|---|---|
| 2:1 MUX | 15% | 20% | 14% |
| 4:1 MUX | 22.43% | 38.12% | 13.99% |
| 8:1 MUX | 25.32% | 41.08% | 16.23% |
| 16:1 MUX | 34.99% | 49.29% | 19.41% |

camouflaging are 26%, 41% and 22%. For 15% camouflaging, the values are 33%, 44% and 29%. From these results, we can observe the linear relation of overhead with respect to the number of camouflaged nets. To further increase the RE effort (discussed next) and to test the flexibility of our proposed mux, we also tested wider i.e., 4:1, 8:1 and 16:1muxes. For these simulations, we only replaced 5% of the nets. Fig. 9(a-c), 10(a-c), and 11(a-c) shows area, delay, and power values of benchmarks. The result of these overhead is shown in Table- 1. From this, we can conclude that wider input muxes can incur affordable increase in design overhead.

We compute RE effort using commercial Automatic Test Pattern Generator (ATPG) as described in [23]. It employs Tetramax ATPG to generate test pattern in order to validate a conjecture about the camouflaged gate. In this work, we employ the same methodology to identify the mux selection. The RE effort is the time taken to set the inputs of the mux to desired values and time taken to propogate the mux output to a primary output (RE effort=$T_{setting}+T_{propogation}$). Each input of the mux has two possible logic (either 1 or 0). For the 2:1 mux we can have two input patterns 2'b10 and 2'b 01 for inputs $I_0$ and $I_1$. We propogate the output to primary output in both cases. If a logic 1 (0) is propagated to the primary output for input pattern 2'b10 then we conclude that input $I_0$ ($I_1$) is selected path. The RE effort includes the time needed by the adversary to validate the guess regarding the mux selection. The ATPG time is determined by executing it on Scientific Linux 6.5 carbon with AMD operton processor (2GHz clock and 32GB RAM). A clock frequency of 1GHz (1ns cycle time) is assumed for each combination of gate-level test pattern generation/application.

Fig. 12(a-b) shows the RE effort for different benchmarks for 2:1 mux using random selection and controllability/ observability based selection. It can be observed that controllability/observability based mux insertion is more effective in increasing the RE effort. Fig 12(c) shows the RE effort for different types of muxes for 5/10/15% camouflaging in c7552 benchmark. It can be clearly observed that 4:1, 8:1 and 16:1 mux could be more effective in increasing the RE effort compared to 2:1 mux while camouflaging same number of nets. This is primarily due to addition of more fake signals in the mux. Since wider mux insertion in logic incurs delay overhead, our next step is to minimize the overhead by targeting off-critical paths. We also intend to experiment with larger benchmarks to validate the effectiveness of interconnect camouflaging in practical circuits. It can be noted that the overheads (area, delay and power) as well as RE effort of the proposed interconnect camouflaging is significantly better than gate camouflaging technique [23].

## V. CONCLUSIONS

We proposed threshold voltage-defined pass transistor based multiplexer to camouflage the interconnects of IPs both logically and physically. Compared to existing split manufacturing, the proposed interconnect camouflaging does not require any process change and does not incur extra assembly cost while increasing the RE effort. Careful selection of nets for camouflaging mitigates the overhead compared to gate camouflaging technique.

## ACKNOWLEDGEMENTS

This work is supposed by Defense Advanced Research Projects Agency under award #D15AP00089